\documentclass[12pt]{iopart}

\usepackage{epstopdf}
\usepackage{hyperref}
\usepackage{graphicx}
\usepackage{rotating}
\usepackage{array}
\usepackage{amssymb}
\usepackage{rotating}
\usepackage{slashed}
\usepackage{cite}

\def\v#1{{\bf#1}}
\def\be{\begin{equation}}
\def\ee{\end{equation}}
\def\bea{\begin{eqnarray}}
\def\eea{\end{eqnarray}}

\newcommand{\bfsigma}{\mbox{\boldmath$\sigma$\unboldmath}}

\def\lcal{\mbox{$\cal L\,$}}

\def\cecal{\mbox{$\cal C$}}
\def\hcal{\mbox{$\cal H$}}
\def\<{\langle}
\def\>{\rangle}

\begin{document}

\title[Dynamical symmetry in a minimal dimeric complex]{Dynamical symmetry in a minimal dimeric complex}

\author{E. Sadurn\'i$^1$ and Y.  Hern\'andez-Espinosa$^2$}

\address{$^1$ Benem\'erita Universidad Aut\'onoma de Puebla, Instituto de F\'isica, Apartado Postal J-48, 72570 Puebla, M\'exico}
\address{$^2$ Universidad Nacional Aut\'onoma de M\'exico, Instituto de F\'isica, Apartado Postal, 04510 Ciudad de M\'exico, M\'exico}

\ead{sadurni@ifuap.buap.mx}

\begin{abstract}

The emergence of non-configurational symmetry is studied in a minimal example. The system under scrutiny consists of a dimeric hexagonal complex with configurational $C_3$ symmetry, formulated as a tight-binding model. An accidental three-fold degeneracy point in parameter space is found; it is shown that an internal $U(3)$ symmetry group operates on Hilbert space, but not on configuration space. The corresponding discrete Wigner functions for the irreducible representations of $C_6 \cong C_3 \times Z_2$ are utilized to show that a $6\times 6$ phase space is sufficient to exhibit an invariant subset. The dynamical symmetry is thus identified with a discrete semi-plane. Some implications on other known hidden symmetries of continuous systems are qualitatively discussed.

\end{abstract}

\maketitle

\section{Introduction}

The concept of dynamical symmetry has been historically relevant for a better understanding of superintegrability in quantum mechanics. The Coulomb problem \cite{pauli_uber_1926, fock_zur_1935, bander_group_1966, stevenson_note_1941} and the isotropic oscillator \cite{bargmann_zur_1936, bargmann_group_1960, bargmann_group_1961, kramer_group_1966, kramer_group_1968} are the paradigms that nature has chosen in the form of atoms and springs, showing us clear realizations of Lie algebraic structures whose dimension exceeds that of physical space. The usefulness of dynamical symmetry, at least in the case of atoms, resides in the explanation of accidental degeneracy, as shown in many textbooks. It also provides a suitable scaffolding for the description of more complex atomic systems around integrability via useful quantum numbers and the transitions between them -- this is the cornerstone of spectroscopy. To the authors' knowledge, there remains an open problem in mathematical physics regarding the reverse statement: that every degeneracy corresponds to a symmetry of a system, regardless of whether its realization is in configuration space or in phase space. Under reasonable assumptions and starting from a prescribed degeneracy of states, one may try to show that the only relevant feature is the existence of a symmetry Lie group, and in this respect the bound states of both the hydrogen atom and the three-dimensional isotropic oscillator have the same SO$(4)$ structure \footnote{The oscillator, of course, has the larger U(3) as symmetry group and sp(6,$\mathbb{R}$) as dynamical algebra.}. Furthermore, the full (Cartan) classification of Lie groups exhausts all other possibilities. But this game is far more complicated when one also recognizes the many forms of degeneracy, either finite (compact group), infinite (non-compact group) or maybe of unkown multiplicity as a function of energy (diophantine problem). Therefore, if a system possesses a set of parameters for which some levels 'accidentally' coalesce, it is difficult in general to show the existence of symmetry in the space of observables (especially if they are canonically conjugate), whereas the trivial answer of an internal symmetry in Hilbert space adds little to our knowledge, despite its correctness.

It is worthwhile to pose these questions in the case of finite systems, where the concept of phase space is attainable if the appropriate Wigner function \cite{wigner_quantum_1932} is employed in the description of the corresponding states. Regardless of the existence of symmetry, such functions can always be put in terms of lattice sites and eigenphases of a certain finite group. This entails the use of two indices that define a two-dimensional phase space where all hidden symmetries can be exposed. In this work we are interested in the spectrum of an electron hopping on a hexagonal polymer made of three dimers, which happens to possess an accidental three-fold degeneracy point. Due to finiteness, we shall be able to answer the question in the previous paragraph without the use of canonical operators $x$ and $p=-i\hbar \partial/\partial x$. Regarding the more general case, the simplicity of our approach will help to {\it build\ }more complex realizations until a continuous limit is reached.  

In connection with the applications of the present work, it is important to mention that dimers are a special realization of q-bits, similar to two-level atoms, provided that each monomer consists of a single-level potential well. The existence of symmetry operators for certain values of the model's parameters can be used to obtain a quantity that remains undisturbed by evolution or other unitary operations. Also, in the control of certain states, a geometric (Berry) phase \cite{berry_quantal_1984} can be identified when the system is driven through a loop. For three-fold degeneracy points, this quantity has been carefully computed \cite{garg_berry_2010, ceulemans_berry_1991, samuel_topological_2001}. More applications related to hexagonal or dimeric structures can be found \cite{montambaux_2009, bittner_observation_2010, barkhofen_disordered_2013, bellec_tight-binding_2013, uehlinger_artificial_2013}; in particular, we have applied three-fold degeneracy to achieve level inversion and negative couplings in a tight-binding chain equipped with a quasi-spin polarizer \cite{rosado_stern-gerlach_2016}. 

Structure of the paper: In section \ref{sec:1} we describe tight-binding polymers with degeneracy, reaching the conclusion that an accidentally degenerate system of three dimers is indeed minimal. We proceed to the diagonalization of a model hamiltonian and a full hamiltonian with pairwise site-to-site coupling, and we describe the geometric configurations containing triplets. The dynamical algebra of the problem is explicitly written. In section \ref{sec:2} we define a $6\times 6$ phase space and find the discrete Wigner function of a triplet for three cases: three-fold accidental degeneracy, two-fold degeneracy with unbroken $C_3$ and no degeneracy with broken time reversal invariance. For the first case, we give a description of the invariant locus in such a phase space. In section \ref{sec:3} we discuss the phase space for a collection of polymers and the emergence of continuous symmetries. 

\section{On polymers and minimal systems \label{sec:1}}       

Our line of reasoning consists in finding the simplest polymeric complexes where degeneracy can be regarded as accidental, i.e. level crossing that is not inherent to polygonal $C_n$ or permutational $S_n$ groups. In addition to this requirement, our main assumption is that our system contains single-level sites and all couplings between sites are positive quantities; this is the case when localized wave functions are real and positive, for the integral overlaps that constitute tunneling amplitudes must have such a quality. A good example is a spherically symmetric ground state of an atom in the vicinity of a positively charged ion; another example is a carbon $\pi$ orbital orthogonal to the molecular plane in benzene or graphene \cite{geim_2007}. Complex couplings are irrelevant in open configurations, while their presence in loops represent magnetic fluxes, but here we deal with systems in the absence of additional external fields, therefore only real positive hopping amplitudes will be considered in this section. 

Our polymers are indeed minimal. The simplest tight-binding models with one, two and three identical dimers are $2 \times 2$, $4 \times 4$ and $6 \times 6$ hamiltonians enjoying the $S_2$ permutational symmetry. The $2 \times 2$ is our building block. The $4 \times 4$ array has no accidental degeneracy in parameter space: if all couplings are equal, a singlet and a triplet are found, but the planar array of four sites can be promoted to a tetrahedron of equal edges, corresponding to a configurational (geometric) symmetry. The $6 \times 6$ system with broken $C_6$ but unbroken $C_3$ symmetry has a crossing point in the spectrum, containing one of the doublets and one of the singlets. The configuration lies somewhere between a star graph and a hexagon, with broken $C_{3V}$ see figure (\ref{fig1}). The corresponding couplings are not identical, therefore no higher-dimensional construction can map the system into a geometrically regular polytope.

Other small non-dimeric systems can be discussed, but once more, their degeneracies are inferred from configuration space. An equilateral trimer has a doublet and a singlet. A polygon with $C_5$ symmetry has two doublets and a singlet, and the existence of a higher (quadruple) degeneracy point corresponds to equal couplings of all elements; the corresponding graph has $S_n$ symmetry and can be embedded in four-dimensional space as a polytope whose projections are regular tetrahedra, which constitute again a symmetry of the configurational type.

\subsection{Our physical system}

\begin{figure}[h!]
\begin{center}
\includegraphics[width=15cm]{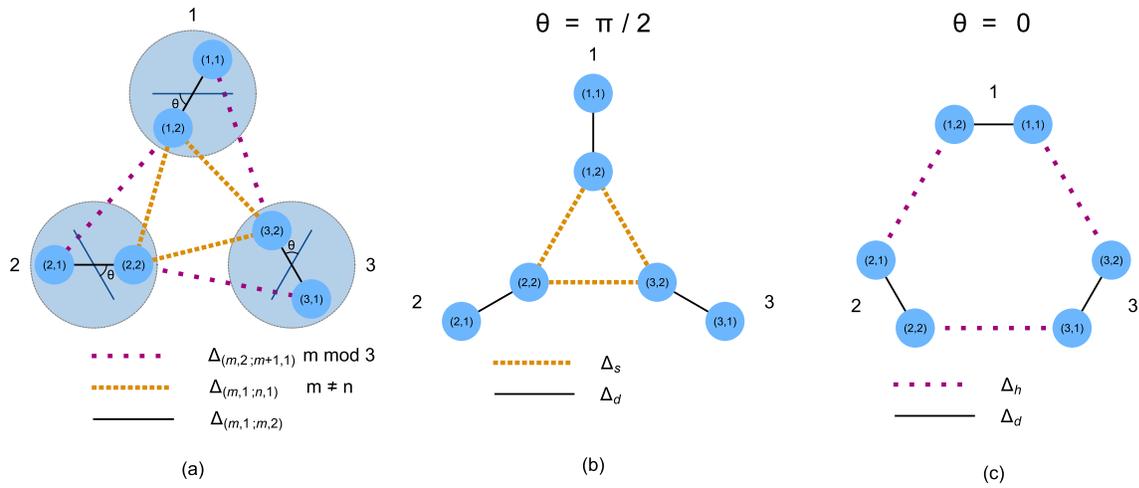}
\caption{\label{fig1} Configuration of dimeric systems with $C_3$ symmetry. Panel (a) shows an arbitrary twisting angle with the corresponding couplings labeled according to intra-dimer and dimer indices. Panel (b) is a star configuration. Panel (c) is a hexagon.}
\end{center}
\end{figure}

Microwave resonators \cite{franco-villafane_first_2013}, bent waveguides with corners \cite{rivera-mocinos_inverse_2016} and, in general, arrays of potential wells such as molecular structures, motivate the introduction of tight-binding Hamiltonians. Consider the hexagon and the star graph in fig. \ref{fig1}. Each point or site is denoted by $(m,n)$ where $m=1,2,3$ is the dimer label and $n=1,2$ is the intra-dimer site. The regular hexagonal configuration with time reversal invariance is known to possess two doublets and two singlets. This remains true if the $C_6$ symmetry is broken (as in the figure), but $C_3$ is preserved: the real positive couplings

\bea
\Delta_{(m,n;l,k)} = \int d^2x \psi_{m,n}(x) H \psi_{l,k}(x), \qquad \psi_{m,n}(x) \equiv \< x | m,n \>
\label{1}
\eea
make $H=H^*$ in this reduced Hilbert space of localized wave functions, so the two conjugate representations of $C_3$ must have eigenfunctions with the same energy (forming thus real wavefunctions by linear combinations). This is the doublet of the equilateral triangle; there is also a singlet. With the additional dimeric structure, we obtain two copies of such levels, but their energies are repelled (or separated) by the intra-dimer coupling. Due to the existence of symmetric and antisymmetric states in each dimer, there can be effectively positive and negative couplings between {\it eigenstates\ }of each dimer and the two copies of each singlet and doublet appear inverted in the energy axis, see fig. \ref{fig2}(a) for small values of $\theta$. The star graph, on the other hand, has two non-inverted copies of the triangle spectrum, but again repelled or displaced with respect to each other due to intra-dimer interaction. One can move continuously from the hexagonal configuration to the star configuration by a rotation of dimers around their centres. From the continuous evolution of the spectrum with respect to the angle $\theta$, one infers that there must be a critical value $\theta_c$ for which the lower singlet and doublet levels have crossed. This critical angle does not correspond to a restoration of $C_{3V}$, as shown by the first panel of figure \ref{fig1}, so the resulting three-fold degeneracy must be 'accidental'. We must prove that this is the outcome of a hidden symmetry in phase space. These general considerations can be substantiated by using a concrete model Hamiltonian:  

\begin{figure}[h!]
\begin{center}
\includegraphics[width=15cm]{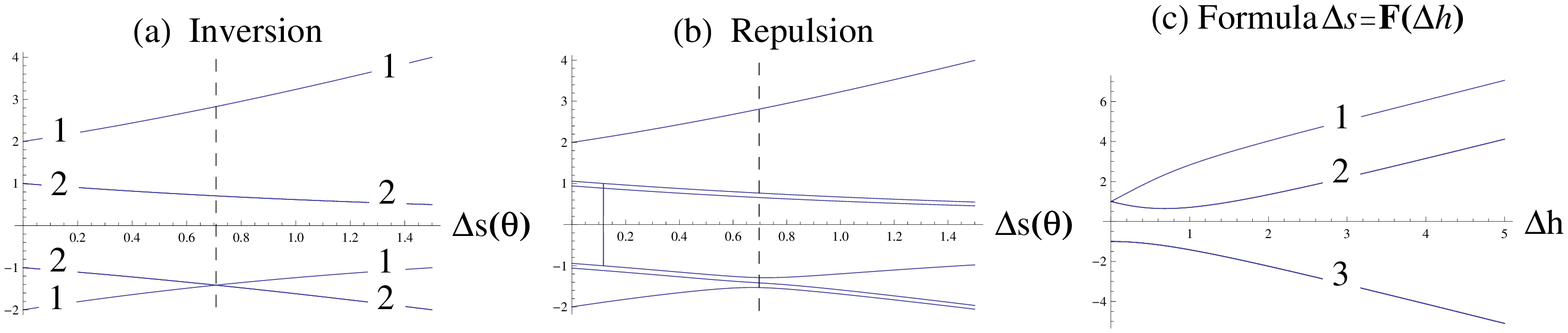}
\caption{\label{fig2} Evolution of energy levels with a geometrical parameter, numbers indicate level degeneracy. Panel (a) shows the level inversion and the crossing point for unbroken $C_3$. Panel (b) shows a level splitting due to explicit $C_3$ symmetry breaking where at least one of the dimer lengths is different. Panel (c) shows the level evolution at critical angle as a function of one of the couplings; this corresponds to a one-parameter family of hamiltonians with triplets. }
\end{center}
\end{figure}

\bea
H =\left( \begin{array}{cccccc} E_0 & \Delta_d & 0&0&0&\Delta_h \\ \Delta_d & E_0 & \Delta_h & \Delta_s & 0& \Delta_s \\
0 & \Delta_h & E_0& \Delta_d & 0& 0 \\ 0 & \Delta_s & \Delta_d & E_0 & \Delta_h & \Delta_s \\ 0&0&0& \Delta_h & E_0 & \Delta_d \\
\Delta_h & \Delta_s & 0 & \Delta_s & \Delta_d & E_0  \end{array}\right)
\label{2}
\eea
where $\Delta_d = \Delta_{(m,1;m,2)}$ are intra-dimer couplings, $\Delta_h = \Delta_{(m,2;n,2)}$ are couplings that favour the hexagonal configuration and $\Delta_s = \Delta_{(m,1;n,1)}$ are couplings at the centre of the array that favour the star configuration. From this model Hamiltonian we can obtain the analytical condition for the existence of triple degeneracy. Let us ignore the on-site energy $E_0$ without loss of generality. First, we diagonalize each dimeric block with a Hadamard matrix

\bea
U= \frac{1}{\sqrt{2}} \left( \begin{array}{cc} +1 & +1 \\ +1 & -1 \end{array}\right), \qquad H^{(1)} = (\v 1_3 \otimes U)^{\dagger} H (\v 1_3 \otimes U).
\label{3}
\eea
Then we gather all lower dimer levels in a $3\times3$ block by means of permutations; similarly for upper dimer levels. The permutation matrix $P$ is such that $H^{(2)}=P^{T} H^{(1)} P$, leading to

\bea
H^{(2)} = \left( \begin{array}{cccccc} \Delta_d &\Delta_+ &\Delta_+ &0& \Delta_- & -\Delta_+  \\  
\Delta_+ & \Delta_d & \Delta_+ & -\Delta_+ & 0& \Delta_-  \\   \Delta_+ &\Delta_+ &\Delta_d &\Delta_- & -\Delta_+ &0 \\
0&-\Delta_+&\Delta_-& -\Delta_d  & -\Delta_-& -\Delta_- \\ \Delta_-&0&-\Delta_+ &-\Delta_-&-\Delta_d&-\Delta_- \\
-\Delta_+& \Delta_-&0 & -\Delta_-&-\Delta_- & -\Delta_d              \end{array}\right),
\label{4}
\eea
with $\Delta_{\pm} \equiv (\Delta_h \pm \Delta_s)/2$. Now we use the $C_3$ basis to diagonalize both $3 \times 3$ blocks in the diagonal of $H^{(2)}$:

\bea
\fl U_{C_3} = \frac{1}{\sqrt{3}}  \left( \begin{array}{ccc} 1 &1 &1  \\ e^{i 2\pi/3 } & e^{i 4\pi/3 } & 1 \\ e^{i 4\pi/3 } & e^{i 2\pi/3 } & 1 \end{array}   \right), \qquad H^{(3)} = (\v 1_2 \otimes U_{C_3})^{\dagger} H^{(2)} \left(  \v 1_2 \otimes U_{C_3}  \right) 
\label{5}
\eea
with the following result
\bea
\fl H^{(3)} =  \left( \begin{array}{cc} X_+ & Y \\ Y^{\dagger} & X_- \end{array}\right), \quad X_{\pm} = \mbox{diag} \left\{ \pm(\Delta_d-  \Delta_{\pm}), \pm(\Delta_d-  \Delta_{\pm}), \pm(\Delta_d + 2  \Delta_{\pm})    \right\} \nonumber \\
Y=\mbox{diag} \left\{ (\Delta_s+ i \sqrt{3} \Delta_h)/2, (\Delta_s- i \sqrt{3} \Delta_h)/2, -\Delta_s    \right\}.
\label{6}
\eea
Finally, we observe that this operator contains uncoupled $2\times2$ blocks, so we diagonalize them to obtain the spectrum:

\bea
E_{\mbox{\scriptsize doublets}}^{\pm} = -\frac{\Delta_s}{2} \pm \sqrt{\left(\Delta_d - \frac{\Delta_h}{2}\right)^2+ \left( \frac{\Delta_s}{2} \right)^2 + 3 \left( \frac{\Delta_h}{2} \right)^2} \\
E_{\mbox{\scriptsize singlets}}^{\pm} = \Delta_s \pm \sqrt{(\Delta_d + \Delta_h)^2+ \Delta_s^2}. 
\label{7}
\eea
Here we see that $E_{\mbox{\scriptsize doublets}}^{-}=E_{\mbox{\scriptsize singlets}}^{-}$ is attainable. After some trivial algebraic steps, this condition results in

\bea
\frac{\Delta_s}{\Delta_d} = \frac{(\Delta_h/\Delta_d)}{\sqrt{1 + (\Delta_h / \Delta_d)^2}} \equiv F(\Delta_h/\Delta_d),
\label{8}
\eea
which constrains the star and hexagonal couplings to a curve, with coordinates $\Delta_s / \Delta_d , \Delta_h / \Delta_d$  normalized with respect to intra-dimer $\Delta_d$. Therefore there is a one-parameter family of Hamiltonians with triple points, and none of its members  contain equal couplings (no permutational symmetry) except for the trivial case  $\Delta_h=\Delta_s=0$. The value of $\theta_c$ must change as a function of the curve's coordinates. According to (\ref{8}), triple points are impossible for $\Delta_s > \Delta_d$,  and $\theta_c$ does not exist. Very weak intra-dimer couplings are an example of this. Sometimes, the localized waves are modeled by exponential tails (e.g. in a cylindrical resonator the Bessel function $K$ can be approximated by decaying exponentials \cite{sadurni_playing_2010}), so the couplings (\ref{1}) also decay exponentially as functions of the separation distance. With the help of (\ref{8}) and trivial geometrical considerations, the critical angle $\theta_c$ can be obtained as a function of inter dimer separations. For dimers that are far apart, the triple degeneracy is removed. Only for tight systems do we see the phenomenon in question, but this does not preclude the existence of a Berry phase for more dilute arrays of dimers where the loop parameterized by $\theta$ does not encounter the triple point. See figure \ref{fig1.1} for distance dependent plots of the degenerate triplet. Panels (a) and (b) are a comparison between model hamiltonian (\ref{2}) and a tight-binding hamiltonian of six sites with couplings between all pairs and modelled by an exponential law that decays with the separation distance.

\begin{figure}[h!]
	\begin{center}
		\includegraphics[width=12cm]{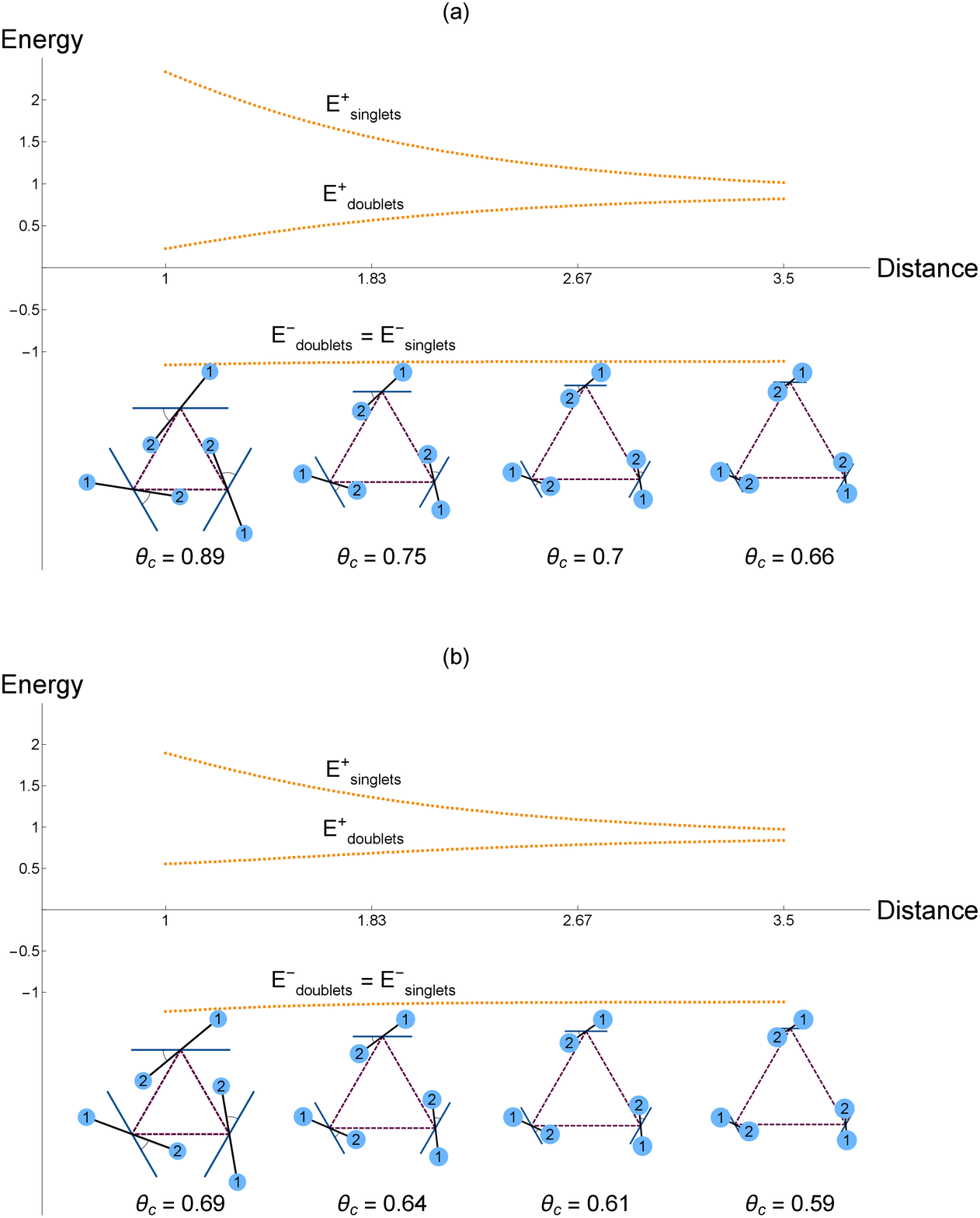}
		\caption{\label{fig1.1} Energy levels for various configurations of distance and angles with triple degeneray $E^- _{\rm doublets} = E^- _ {\rm singlet}$. (a) Eigenstates of a tight-binding hamiltonian with all-site pairwise interactions and (b) eigenstates of the model hamiltonian (\ref{2}). $\theta$ is given in radians, and the distance is given in units of dimeric length.}
	\end{center}
\end{figure}

\subsection{Dynamical algebra}

As in any finite system, the Hamiltonian can be put in terms of SU$(N)$ hermitian generators. Here $N=6$, but a close inspection of (\ref{2}) makes us consider the direct product of irreducible representations $(1/2,1)$ of SU$(2) \otimes $SU$(2)$. Moreover, if the $C_3$ symmetry is unbroken, the second SU$(2)$ factor can be replaced by the cyclic group generated by a $2\pi/3$ rotation of the polymer plane. Let us show this explicitly: We have vector couplings of the type

\bea
H= v_0 + \v v \cdot \bfsigma 
\label{9}
\eea
with
\bea
v_0 = \frac{\Delta_s}{2} \left( J_+ + J_- + J_+^2 + J_-^2 \right) \\
v_1 =  \Delta_d + \Delta_h \left( J_+ + J_- + J_+^2 + J_-^2 \right) \nonumber \\
v_2 = i \Delta_h \left( J_+^2 + J_- - J_-^2 - J_+ \right) \nonumber \\
v_3 = - \frac{\Delta_s}{2} \left( J_+ + J_- + J_+^2 + J_-^2 \right) \nonumber
\label{9.1}
\eea
and the algebraic relations
\bea
\left[ J_+, J_- \right] = 2 J_3, \quad J_+^3=J_-^3=0, \quad J_3 = \mbox{diag} \left\{1,0,-1 \right\}, \\
\left[ \sigma_+, \sigma_- \right] = 2 \sigma_3, \quad \sigma_+^2=\sigma_-^2=0, \quad \sigma_3 = \mbox{diag} \left\{1,-1 \right\}, \nonumber \\
\left[ J_{\pm}, \sigma_{\pm} \right] = \left[ J_{\pm}, \sigma_{3} \right] = \left[ J_{3}, \sigma_{\pm} \right]=\left[ J_{3}, \sigma_{3} \right]=0. \nonumber
\label{10}
\eea
Because of the vanishing powers $J_{\pm}^3=0$, a finite rotation operator $T=J_+ + J_-^2$ can be introduced

\bea
\fl v_0 = -v_3 = \frac{\Delta_s}{2} \left( T  + T^{\dagger} \right), \quad v_1 = \Delta_d + \Delta_h (T  + T^{\dagger}), \quad v_2 = i \Delta_h ( T^{\dagger}-T), \nonumber \\
\fl T^3=T, \quad \left[T  , T^{\dagger}\right]=0,
\label{11}
\eea
which further reduces the algebra to $\left[ v_{\lambda}, \sigma_j \right]=\left[ v_{\lambda}, v_{\mu} \right]=0, \, \forall \, \lambda, \mu, j$. The accidental degeneracy is produced by $\left[ H, I \right]=0$ if

\bea
\fl I =  \alpha | E_{\mbox{\scriptsize doublet}}^{-} \>\< E_{\mbox{\scriptsize singlet}}^{-}| + \beta | E_{\mbox{\scriptsize doublet}}^{-} \>\< E_{\mbox{\scriptsize doublet}}^{+}| + 
\gamma | E_{\mbox{\scriptsize doublet}}^{+} \>\< E_{\mbox{\scriptsize singlet}}^{-}|+\mbox{h.c.}
\label{12}
\eea
for non-zero coefficients and only when (\ref{8}) holds. Here, the freedom of complex parameters $\alpha$, $\beta$ and $\gamma$ shows an internal six-dimensional manifold contained in the nine-dimensional Lie group U(3) as symmetry, but this is only an illusion in Hilbert space. The symmetry must be defined in phase space, as we discuss in the following.  

\section{Phase space in a $6\times 6$ grid \label{sec:2}}

In our system, the $C_6$ symmetry is never fulfilled, but our aim is to use the eigenstates of $C_6 \cong C_3 \times Z_2$ as momentum marks in the vertical axis, conjugate to the six positions $(m,n)$ in the horizontal axis. One can pass from one axis to the other by linear combinations analogous to Fourier transforms. For the $C_3$ generator we have $T  | m, n \> = | (m + 1) , n \>$ where $(m)\equiv m \, \mbox{\small mod}3$ \footnote{It what follows, we use $1,2,3$ excluding $0$}.  The transformed states are denoted by a subindex 1:

\bea
\fl | k, n \>_{1} = \frac{1}{\sqrt{3}} \sum_{q=1,2,3} e^{2\pi i q k/3} | q, n \>, \quad T | k, n \>_{1} = e^{-2 \pi i k/3}| k, n \>_{1}, \quad k=1,2,3.
\label{13}
\eea
For the $Z_2$ part, we have the obvious symmetric and antisymmetric combinations, denoted by the subindex 2:

\bea
|q, s \>_2 = \frac{1}{\sqrt{2}} \left(  (-1)^s | q,1  \>  + | q, 2 \>  \right), \quad s= 1, 2.
\label{14}
\eea
The full transformation is defined then as

\bea
|k,s \>_{1,2} = \frac{1}{\sqrt{6}} \sum_{q=1,2,3; r=1,2} e^{2\pi i q k/3} (-1)^{r s} | q, r \>.
\label{15}
\eea
Because of unitarity in (\ref{3}) and (\ref{5}), the inverse is well defined. The wavefunctions are denoted by $\psi_{(q),n} = \< q, n  | \psi \>$, $\tilde \psi_{k,s}=_{1,2}\< k, s  | \psi \>$.

In order to define a {\it bona fide\ }Wigner function, it is necessary to analyze each factor in $C_6$ separately, instead of using its eigenphases $1^{1/6}$ directly; this is related to the fact that $6$ is not a prime. The factor $C_3$ allows a simple treatment in terms of $1^{1/3}$ and the corresponding 3 eigenstates of $T$; however $Z_2$ needs at least three summands (not two) in the Wigner function in order to recover the correct marginal distributions.

\subsection{Wigner function for $Z_2$}

There is a covariant definition provided in \cite{klimov_generalized_2017} for any spin, in particular $s=1/2$. For two-level systems one may also resort to \cite{dowling_wigner_1994}. Here we are not interested in marginal distributions over the three possible axes $\sigma_x, \sigma_y, \sigma_z$, but only on the two eigenphases $\pm 1$ of $Z_2$. Our phase space is two-dimensional. The most general function $W(l,\beta)$ of variables $l =1,2, \beta=1,2$ that yields the correct marginals is

\bea
W(1,1)= A |\psi_{1}|^2 + B |\psi_2|^2 + C \Re\left\{ \psi_1 \psi_2^* \right\}, \nonumber \\ 
W(2,1)= (1-A) |\psi_{1}|^2 - B |\psi_2|^2 - C \Re\left\{ \psi_1 \psi_2^* \right\}, \nonumber \\
W(1,2)= (1/2-A) |\psi_{1}|^2 + (1/2-B) |\psi_2|^2 + (1/2-C) \Re\left\{ \psi_1 \psi_2^* \right\}, \nonumber \\
W(2,2)=(A-1/2) |\psi_{1}|^2 + (B+1/2) |\psi_2|^2 + (C-1/2) \Re\left\{ \psi_1 \psi_2^* \right\},
\label{16}
\eea
where $A,B,C$ are arbitrary real numbers and $\Re\left\{ c \right\}\equiv c +c^*$. These relations satisfy $\sum_l W(l,\beta) = |\psi_{\beta}|^2$, $\sum_{\beta} W(l,\beta) = |\tilde \psi_{l}|^2$. They can also be written as

\bea
W(l,\beta) = \sum_{a=1,2; b=1,2} C_{a,b}^{l,\beta} \psi_{a} \psi_b^{*}.
\label{17}
\eea
For a very special choice $A=1/2, B=0, C=1/4$, the non-zero coefficients $C_{a,b}^{l,\beta}$ can be put explicitly in terms of $l,\beta$, so we may use

\bea
C_{1,1}^{l,\beta}= \frac{1+(-)^l}{4}, \quad C_{2,2}^{l,\beta}= \frac{1-(-)^l}{4}, \quad C_{1,2}^{l,\beta}= C_{2,1}^{l,\beta}=\frac{(-)^{\beta}}{4}.
\label{18}
\eea 

\subsection{Wigner function for $C_3$}

Here we proceed directly with the eigenphases

\bea
\fl W_{C_3}(q',k) = \frac{1}{3} \sum_{q=1,2,3} \psi_{(q)} \psi_{(q'-q)}^{*} \exp \left[ i \frac{2\pi k}{3} (2q-q' ) \right], \quad (q'-q)\equiv q'-q \, \mbox{\small mod} 3.
\label{19} 
\eea
It is confirmed that

\bea
\sum_{q'=1,2,3} W_{C_3}(q',k) = \left|  \frac{1}{\sqrt{3}}\sum_{a=1,2,3} \psi_a e^{i 2\pi a k /3} \right| ^2
\label{20}
\eea
and
\bea
\sum_{k=1,2,3} W_{C_3}(q',k) = \left|  \psi_{ (-q') } \right| ^2
\label{21}
\eea
or, equivalently
\bea
\sum_{k=1,2,3} W_{C_3}( -q' \, \mbox{\small mod}3 ,k) = \left|  \psi_{ q' } \right| ^2.
\label{22}
\eea

\subsection{Full Wigner function}

The aim now is to give a full definition for $C_6$. Using our wavefunctions with two indices $(m,n)$, we recover the correct marginal distributions with the definition

\bea
\fl W(l_1,l_2; \beta_1, \beta_2) = \sum_{a=1,2; b=1,2} \sum_{q=1,2,3} \psi_{a,(q)} \psi_{b,(l_2-q)}^{*} C_{a,b}^{l_1,\beta_1} \exp \left[ i \frac{2\pi \beta_2}{3} (2q-l_2) \right].
\label{23}
\eea
\begin{figure}[h!]
\begin{center}
\includegraphics[width=10cm]{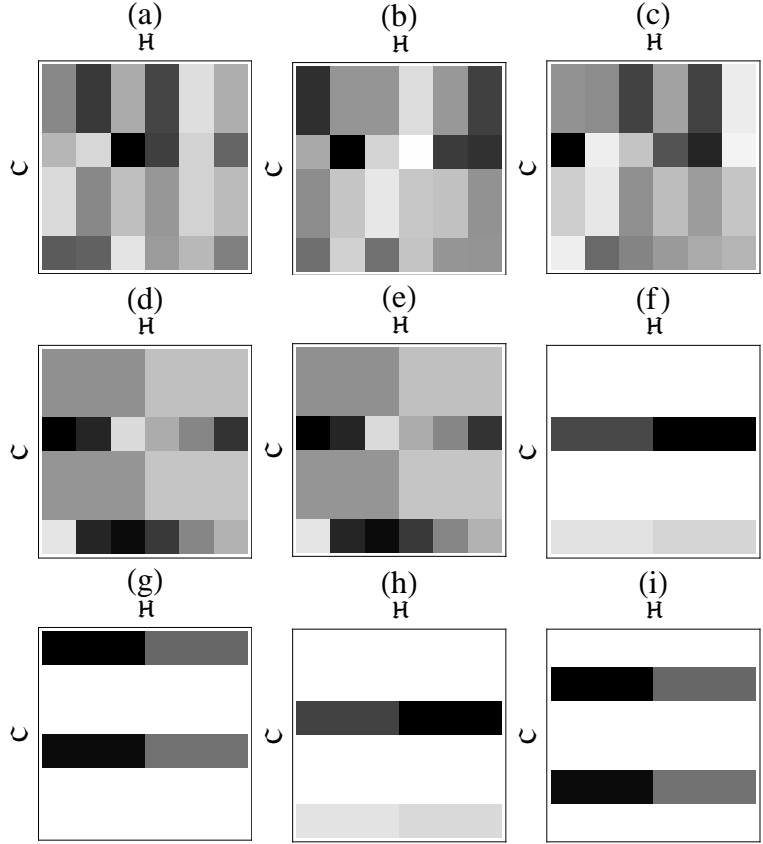}
\caption{\label{fig3} Discrete phase space of triplets. First row: Critical angle with time reversal symmetry, three-fold degeneracy. Arbitrary linear combinations of panels (a), (b), (c) reveal an invariant structure of at least three rows of pixels. Second row: a doublet (d) (e) and a singlet (f) obtained from an increase in the star coupling $\Delta_s$. Third row: Time reversal symmetry breaking with a complex coupling (magnetic flux) showing that only $C_3$ eigenfunctions are allowed. $\hcal$ is the Hilbert space of on-site functions and $\cecal$ its finite Fourier transform.}
\end{center}
\end{figure}
The wavefunctions of interest are contained in the degenerate triplet of (\ref{2}) under the condition (\ref{8}). We study the behaviour of (\ref{23}) for these three solutions in various regimes: 
\newpage
\begin{itemize}
\item[i)] Critical angle $\theta=\theta_c$ with time reversal symmetry. 
\item[ii)] A slightly perturbed configuration away from the triple point, i.e. $F(\Delta_s / \Delta_d) \mapsto F(\Delta_s / \Delta_d) + \delta$. 
\item[iii)] Critical angle with a small magnetic field applied with the minimal substitution $\Delta_h \mapsto \Delta_h e^{i \Phi}$. 
\end{itemize}
Let us denote by $\hcal$ the Hilbert space of localized states labeled by site numbers; the elements of this basis can be regarded as positions, hence $\hcal$ can be associated with configuration space $X$. Similarly, let $\cecal$ be the space of phase eigenstates, which corresponds to $\hcal$ under finite Fourier transforms; $\cecal$ can be associated with momentum space $P$. We adopt the following order: our $6\times 6$ grid is made of rows labeled in top-to-bottom order as $(k,q)=(1,1),(2,1),(3,1),(1,2),(2,2),(3,2)$, corresponding to the $C_3 \times Z_2$ eigenphases. The columns are polymer sites in the order $(m,n)=(1,1),(2,1),(3,1),(1,2),(2,2),(3,2)$. In each row of fig. \ref{3} we show three phase space portraits of the triplet. The first row (a), (b), (c) corresponds to $\theta=\theta_c, \Delta_h/\Delta_d=1.9$ and arbitrary linear combinations of (orthogonal) degenerate states, displaying full occupation of the grid.  The second row shows $\theta\neq \theta_c, \Delta_s = F(1.9) + 0.1$ and a splitting into a doublet (d), (e) and a singlet (f), where the latter displays only two fringes of occupied pixels. The third row (g), (h), (i) is the triplet in the eigenphase basis, where the supports of the $W$ functions do not intersect each other; the parameters of this last set of portraits are remarkably similar to the case $\theta=\theta_c, \Delta_h/\Delta_d=1.9$ but with a slight time reversal symmetry breaking $\Phi=10^{-9} \times \pi/2$ for which the two conjugate representations of $C_3$ must have different energies, albeit very close to each other $\Delta E \sim 10^{-11}$. The small splitting has a strong effect on phase space portraits, as can be verified by comparing the first and third rows of fig. \ref{3}. 

The emergence of hidden symmetry is now evident: By comparing the first and third rows of fig. \ref{3}, we see that all linear combinations of the triplet will reveal full occupation of phase space in various configurations with the same energy. The reason we only consider the lower semi plane is that the upper semi plane is completely correlated to the former, as can be noted by combining (g), (h), (i) with various coefficients. It is worthwhile mentioning that the portraits (d), (e) can be obtained also by combinations of (g), (i). In this case, the resulting rows $(3,1), (3,2)$ of (d), (e) are not empty due to interference terms (cross terms) in $W$ that can be regarded as discrete versions of {\it the cat's smile\ }\cite{atakishiyev_1998}.

As a conclusion, the locus of symmetry is the lower discrete semiplane in phase space, instead of the nine-dimensional Lie manifold U(3).

\section{The emergence of continuous symmetries \label{sec:3}}

\begin{figure}[h!]
\begin{center}
\includegraphics[width=12cm]{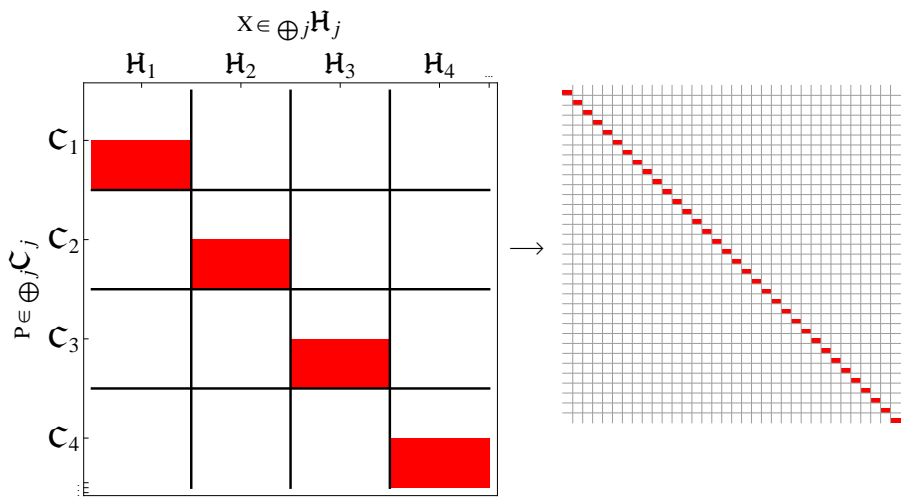}
\caption{\label{fig4} Full phase space for a single particle in a collection of polymers. A direct sum of Hilbert spaces represent position (abscissa) and its finite Fourier transform with $C_6$ eigenphases represents momentum (ordinate). In the right panel we show the limit process towards a continuum.}
\end{center}
\end{figure}

We would like to give some conjectural remarks on the necessary conditions for a continuous symmetry to emerge (more rigorous proofs may be concocted using appropriate measures and metrics). We shall formulate our result in terms of degenerecies. Let us consider a larger system built by a collection of hexamers on which our single particle can jump. The total Hilbert space of localized states is then $\hcal_{\rm tot} = \bigoplus_i \hcal_i$ --a sum, not a product-- while its Fourier transform is the direct sum of eigenphase states $\cecal_{\rm tot} = \bigoplus_j \cecal_j$, supplemented by the eigenphase states of an additional transformation that hops from polymer $i$ to polymer $i+1$. We identify the space of position and momentum states of the full system as $X \leftrightarrow \hcal_{\rm tot}$, $P \leftrightarrow \cecal_{\rm tot}$. A specific value of the position, denoted by X, is identified with a localized eigenstate in the space $X$, similarly for momentum P and space $P$, see figure \ref{fig4}. Using our previous results, the diagonal  blocks $\hcal_i \times \cecal_i$ possess invariant subsets (red boxes) that can be juxtaposed to form a continuous infinite set. This is possible when the number of copies $N \rightarrow \infty$ and the number of polymers configured in critical angles is $g \rightarrow \infty$. We fix their ratio $g/N \rightarrow \lcal$ with $\lcal$ a finite length. If the invariant set of the full system remains discrete, we shall have $\lcal \rightarrow 0$ and if the invariant set is unbounded, we shall have $\lcal \rightarrow \infty$, which necessitates a non-compact group. We focus on compact groups, therefore $0<\lcal < \infty$ as a minimal requirement. The total number of quanta for a degenerate energy is $3g$ and the action can be written as $S=3 \hbar g = 3\hbar N \lcal$. With this, we venture the following statement for ensembles of discrete systems:

\begin{itemize} 
\item[1.] A hidden symmetry is continuous and one-dimensional if the associated degeneracy is proportional to the number of quanta, i.e. the action. 
\item[2.] A system in $d>1$ continuous variables built from an ensemble is integrable if $g \sim N^{d-2}$ and superintegrable if $g \sim N^{p}$ with $p \geq d-1$.
\end{itemize}
The converse statement is evidently false, since integrable hamiltonians can be written in terms of integrals of the motion in arbitrary combinations, with the possibility of removing degeneracy. In our case, it is remarkable that traditional examples of integrable and superintegrable systems follow a pattern consistent with 1 and 2:
\\
\\
{\it The Coulomb problem.\ } The bound states of the non-relativistic hydrogen atom obey $l_{\rm max}=n-1$, $g = (2(n-1) +1)\times n = n(n-1) \sim n^2 $, for the $n$-th level, ignoring a trivial factor of 2 from spin. The system is superintegrable with a group that contains SO(4) $\cong$ SO(3)$\times$SO(3).
\\
\\
{\it The isotropic harmonic oscillator.\ } The group in the previous example solves this problem in $3d$ as well. Moshinsky showed a larger U(3) using ladder operators \cite{moshinsky_book_1996} and $g\sim N^2$. In $2d$ one has U(2), the system is also superintegrable and $g\sim N$. 
\\
\\
{\it Two commensurate oscillators.\ } The degeneracy grows linearly with $N$ for all solutions of $pn+qm = N$, with $n,m$ the oscillator quanta and $p,q \in \mathbb{N}$ relatively primes. The subtleties of the symmetry group U(2) can be found in \cite{draayer_1989}.
\\
\\
{\it The Dirac-Coulomb problem.\ } The system enjoys a supersymmetry that allows to find the bound spectrum exactly. However, the degeneracy is linear in the number of quanta: only the angular momentum projections and the $l, l+1$ orbitals are degenerate (without Lamb shift). The Lippmann-Johnson operator  \cite{dahl_jens_peder_dirackepler_1995, c_v_sukumar_supersymmetry_1985} is a symmetry generator, in addition to the SO(3) group and the Dirac (spin-orbit) operator $K$, but its presence only helps to sustain an additional two-fold degeneracy: In fact, its square can be put in terms of $K^2$ and $H^2$ \cite{katsura_2006} eq. (5, 6). Any possible hidden symmetry must be then discrete, and should be accommodated in bi-spinorial degrees of freedom. The explicit invariant sets are not reported in the literature, but a Wigner function is proposed in \cite{rabitz_2016}.
\\
\\
{\it The Landau electron.\ } This problem has infinite degeneracy. The presence of only one chiral harmonic excitation number (and not two) in the hamiltonian reveals commutability with the Heisenberg algebra, i.e. non-compact group, in addition to cylindrical symmetry.
\\
\\
{\it The Dirac oscillator.\ } This problem has infinite degeneracy \cite{moshinsky_1989}, but the symmetry algebra can be decomposed in compact and non-compact subalgebras   \cite{quesne_1990}. The compact part is so(4) and the total degeneracy is quadratic, made from a linear factor coming from the sequence $(N\pm1,j\mp1), (N\pm2,j\mp2), ...$ and another linear factor from the $2j+1$ projections. The system is superintegrable. For related spin-orbit systems, see \cite{winternitz_2006}.
\\
\\
{\it Circular and square cavities, separability.\ } Diophantine equations offer a challenge for the computation of degeneracies. Without delving into Hilbert's tenth problem, it suffices to say that degeneracies $g=1,2$ occur irregularly for a square box, and are at least two-fold $\pm m$ for a circular shape. Since $d=2$, our statement 2 is not contradicted. Moreover, at high energies, the formula $n_x^2 + n_y^2 = E$ is that of a circle, so the square box has approximate linear degeneracy for large $E$. Similarly, the circular shape of radius $R$ and energy $E=k^2/2$ obeys $J_m(kR)=0$; using the asymptotic form of the Bessel function, the energy equation $\cos \left[ kR- \pi(2m+1)/4 \right]=0$ is solved by $kR = \pi(2m+4n+3)/4$ with $m,n \geq 0$, and the combination $2n+m$ also has linear degeneracy. Indeed, it is a simple exercise to put the classical energy in terms of two action variables for this problem, defining a family of constant energy curves in the plane of actions. Despite of this, the two classical problems are not recognized in the literature as superintegrable.

\section{Conclusions}

We have found a triple degeneracy point in a system consisting of three dimers. The resulting geometry does not correspond to a recognizable configurational symmetry. It was argued that polymers with a lesser number of sites do not show this phenomenon; in this sense, the system is regarded as minimal. In order to understand the nature of hidden symmetry, we defined a finite phase space and studied the behaviour of the corresponding Wigner function at critical angle and under two types of symmetry breaking: time reversal and polygonal. Since this could be done in the simplest possible case, we went further and built a collection of these objects with the purpose of explaining hidden continuous symmetries starting from degeneracies. Some paradigmatic systems follow our pattern.

As an outlook, we envisage a two-dimensional periodic construction made of critically configured polymers, with the aim of emulating the emergence of purely geometric magnetic fields, with applications to transport properties of electromagnetic waves and the artificial realization of the (non-anomalous) quantum Hall effect without charge carriers \cite{eladio_2015}. From the point of view of anomalous spectral statistics in $C_3$ geometries \cite{leyvraz_1996} and false T violation \cite{berry_1986}, this construction would be plausible and desirable.

\section*{References}
\bibliographystyle{iopart-num}

\begin{thebibliography}{10}
\expandafter\ifx\csname url\endcsname\relax
  \def\url#1{{\tt #1}}\fi
\expandafter\ifx\csname urlprefix\endcsname\relax\def\urlprefix{URL }\fi
\providecommand{\eprint}[2][]{\url{#2}}



\bibitem{stevenson_note_1941}
Stevenson A~F 1941 {\em Phys. Rev.\/} {\bf 59} 842--843

\bibitem{bander_group_1966}
Bander M and Itzykson C 1966 {\em Rev. Mod. Phys.\/} {\bf 38} 330--345


\bibitem{fock_zur_1935}
Fock V 1935 {\em Z. Phys.\/} {\bf 98} 145--154

\bibitem{pauli_uber_1926}
Pauli W 1926 {\em Z. Phys.\/} {\bf 36} 336--363

\bibitem{bargmann_group_1960}
Bargmann V and Moshinsky M 1960 {\em Nucl. Phys.\/} {\bf 18} 697

\bibitem{bargmann_zur_1936}
Bargmann V 1936 {\em Z. Phys.\/} {\bf 99} 576--582

\bibitem{bargmann_group_1961}
Bargmann V and Moshinsky M 1961 {\em Nucl. Phys.\/} {\bf 23} 177--199

\bibitem{kramer_group_1966}
Kramer P and Moshinsky M 1966 {\em Nucl. Phys.\/} {\bf 82} 241--274

\bibitem{kramer_group_1968}
Kramer P and Moshinsky M 1968 {\em Nucl. Phys. A\/} {\bf 107} 481--522

\bibitem{wigner_quantum_1932}
Wigner E 1932 {\em Phys. Rev.\/} {\bf 40} 749--759

\bibitem{berry_quantal_1984}
Berry M~V 1984 {\em Proc. R. Soc. London, Ser. A\/} {\bf 392} 45--57


\bibitem{garg_berry_2010}
Garg A 2010 {\em Am. J. Phys.\/} {\bf 78} 661--670

\bibitem{samuel_topological_2001}
Samuel J and Dhar A 2001 {\em Phys. Rev. Lett.\/} {\bf 87}


\bibitem{ceulemans_berry_1991}
Ceulemans A and Szopa M 1991 {\em J. Phys. A\/} {\bf 24} 4495--4509

\bibitem{montambaux_2009}
Montambaux G and Piechon F and Fuchs J N and Goerbig M O 2009 {\em Phys. Rev. B\/} {\bf 80} 153412


\bibitem{uehlinger_artificial_2013}
Uehlinger T, Jotzu G, Messer M, Greif D, Hofstetter W, Bissbort U and Esslinger
  T 2013 {\em Phys. Rev. Lett.\/} {\bf 111} 185307

\bibitem{barkhofen_disordered_2013}
Barkhofen S, Bellec M, Kuhl U and Mortessagne F 2013 {\em Phys. Rev. B\/} {\bf
  87} 035101

\bibitem{bittner_observation_2010}
Bittner S, Dietz B, Miski-Oglu M, Oria~Iriarte P, Richter A and Sch\"afer F 2010
  {\em Phys. Rev. B\/} {\bf 82} 014301

\bibitem{bellec_tight-binding_2013}
Bellec M, Kuhl U, Montambaux G and Mortessagne F 2013 {\em Phys. Rev. B\/} {\bf
  88} 115437

\bibitem{rosado_stern-gerlach_2016}
Rosado A~S, Franco-Villafa\~ne J~A, Pineda C and Sadurní E 2016 {\em Phys. Rev.
  B\/} {\bf 94} 045129

\bibitem{geim_2007}
Geim A and Novoselov K S 2007 {\em Nature Materials\/} {\bf 6} 183–191




\bibitem{franco-villafane_first_2013}
Franco-Villafa\~ne J~A, Sadurn\'i E, Barkhofen S, Kuhl U, Mortessagne F and
  Seligman T~H 2013 {\em Phys. Rev. Lett.\/} {\bf 111} 170405

\bibitem{rivera-mocinos_inverse_2016}
Rivera-Moci\~nos E and Sadurn\'i E 2016 {\em J. Phys. A\/} {\bf 49} 175302


\bibitem{sadurni_playing_2010}
Sadurn\'i E, Seligman T~H and Mortessagne F 2010 {\em New J. Phys.\/} {\bf 12}
  053014

\bibitem{klimov_generalized_2017}
Klimov A~B, Romero J~L and de~Guise H 2017 {\em J. Phys. A\/} {\bf 50} 323001





\bibitem{dowling_wigner_1994}
Dowling J~P, Agarwal G~S and Schleich W~P 1994 {\em Phys. Rev. A\/} {\bf 49}
  4101--4109


\bibitem{atakishiyev_1998}
Atakishiyev N M and Chumakov S M and Wolf K B 1998 {\em J. Math. Phys.\/} {\bf 39} 6247

\bibitem{moshinsky_book_1996}
Moshinsky M and Smirnov Y F 1996 {\em The Harmonic Oscillator in Modern Physics \/} 1st ed
(Contemporary Concepts in Physics vol 9) (The Netherlands: Hardwood Academic Publishers)
ISBN 3718606208

\bibitem{draayer_1989}
Rosensteel G and Draayer J P 1989 {\em J. Phys. A: Math. Gen.\/} {\bf 22} 1323






\bibitem{dahl_jens_peder_dirackepler_1995}
{Dahl Jens Peder} and {J{\o}orgensen Thomas} 1995 {\em Int. J. Quantum Chem.\/}
  {\bf 53} 161--181 \urlprefix\url{https://doi.org/10.1002/qua.560530204}

\bibitem{c_v_sukumar_supersymmetry_1985}
{C V Sukumar} 1985 {\em J. Phy. A\/} {\bf 18} L697 ISSN 0305-4470
  \urlprefix\url{http://stacks.iop.org/0305-4470/18/i=12/a=002}

\bibitem{katsura_2006}
Katsura H and Aoki H 2006 {\em J. Math. Phys.\/} {\bf 47} 032301

\bibitem{rabitz_2016}
Cabrera R and Campos A G and Bondar D I and Rabitz H A 2016 {\em J. Phys. A: Math. Gen.\/} {\bf 94} 052111


\bibitem{moshinsky_1989}
Moshinsky M and Szczepaniak A 1989 {\em J. Phys. A: Math. Gen.\/} {\bf 22} L817

\bibitem{quesne_1990}
Quesne C and Moshinsky M 1990 {\em J. Phys. A: Math. Gen.\/} {\bf 23} 2263

\bibitem{winternitz_2006}
Winternitz P and Yardusen I 2006 {\em J. Math. Phys.\/} {\bf 47} 103509

\bibitem{eladio_2015}
Sadurn\'i E and Rivera-Moci\~nos E 2015 {\em J. Phys. A: Math. Gen.\/} {\bf 48} 405301

\bibitem{leyvraz_1996}
Leyvraz F and Schmidt C and Seligman T H 1996 {\em J. Phys. A: Math. Gen.\/} 29 L875

\bibitem{berry_1986}
Robnik M and Berry M V 1986 {\em J. Phys. A: Math. Gen.\/} {\bf 19} 669


\end{thebibliography}

\providecommand{\newblock}{}

\end{document}